\documentclass[prd,aps,preprint,epsfig,floats,nofootinbib]{revtex4}
\usepackage[dvips]{epsfig}
\usepackage{graphicx}
\usepackage{amsmath}

\input{axodraw.sty}
\input{epsf.sty}
\begin{document}

\preprint{\vbox{
\hbox{UMD-PP-05-051} }}


\title{{\Large\bf Some Implications of
Neutron Mirror Neutron Oscillation }}

\author{\bf  R.N. Mohapatra$^1$, S. Nasri$^1$ and
S.  Nussinov $^{1,2}$ }

\address{$^1$ Department of Physics and Center for String and Particle 
Theory, University of Maryland, College Park,
MD 20742, USA \\
$^2$  Sackler Faculty of Science, Tel Aviv University,
 Tel Aviv, Israel}

\date{August , 2005}
\begin{abstract}
We comment on a recently discussed possibility of oscillations
between neutrons and degenerate mirror neutrons in the context of
mirror models for particles and forces. It has been noted by Bento
and Berezhiani that if these oscillations occurred at a rate of
$\tau^{-1}_{NN'}\sim sec^{-1}$, it would help explain putative
super GKZ cosmic ray events provided the temperature of the mirror
radiation is $\sim 0.3-0.4$ times that of familiar cosmic
microwave background radiation. We discuss how such oscillation
time scales can be realized in mirror models and find that the
simplest nonsupersymmetric model for this idea requires the
existence of a low mass ($30-3000$ GeV) color triplet scalar or
vector boson.  A supersymmetric model, where this constraint
can be avoided is severely constrained by the requirement of
 maintaining a cooler mirror sector. We also find that the 
reheat temperature after
inflation in generic models that give fast $n-n'$ oscillation
 be less than about 300 GeV in order to maintain the required relative 
coolness of the mirror sector.
 \end{abstract}

\vskip1.0in

\maketitle

Oscillations of electrically neutral particles such as the $K^0$
(to $\bar{K}^0$), $B_d$ (to $\bar{B}_d$) mesons and the neutrinos
have provided us with a wealth of information about the nature of
fundamental interactions. If baryon number is not a good symmetry
of nature (as is implied by grand unified theories), other
oscillation phenomena can take place among electrically neutral
systems e.g. neutron-anti-neutron oscillation\cite{moh} and
hydrogen-anti-hydrogen oscillation\cite{FGS, MSAW}. In the context
of mirror universe models\cite{mirror,mirror1,mirror2} where
 there is an identical copy of both the forces and matter
of the standard model present side by side with the known forces
and matter, a new possibility arises whereby a neutron ($n$) can
oscillate into its mirror partner $n'$. Since the mirror
particles are supposed to have only gravitational or similarly
suppressed interactions, the mirror neutron will not be detected
by our measuring devices and the process $n-n'$ will cause the
disappearance
of the neutron. Clearly, such an $n-n'$ oscillation will have
phenomenological and astrophysical implications.

 Recently Berezhiani and Bento \cite{BB}  have pointed out
 that all observations are consistent with the possibility
that neutrons can oscillate to mirror neutrons on a very short
time scale of about one second and that $n-n'$
oscillations with this rate can allow some Super GZK\cite{GKZ} ultra high 
energy (UHE) cosmic
rays to reach us providing that the cosmic microwave background
temperature in the mirror sector is significantly lower, by say
$x=T'/T \sim 1/3$ than in our visible sector.
The paper\cite{BB} correctly pointed out that :

 i) the strict bounds\cite{sno} on $n
\longrightarrow 3 \nu$ utilizing the resultant nuclear excitation
and gamma emission do not apply here as $n \rightarrow n'$
oscillations inside nuclei is forbidden by energy conservation;

ii) the Grenoble bounds \cite{GR} on the neutron
-anti-neutron oscillation rate $\tau_{n-\bar{n}}^{-1} < 10^{-8}
sec ^{-1}$ coming from the absence of dramatic anti-neutron
annihilation events also does not apply to $n \rightarrow n'$
oscillation .

An $n-n'$ oscillation rate as large as $1 sec^{-1}$
would of course imply a reduction of the reactor neutron flux by as much
as 1\%. In view of other neutron loss mechanisms  at this level which can 
amount to as much as 5\% or more, $1$ sec$^{-1}$. 
can be used as a crude
upper bound on the rate and is presumably not inconsistent with
observations.

Coming to galactic propagation of neutrons, first point to note is that 
since the Lorentz enhancement of the magnetic field compensates
for time dilation, in order to have unquenched $n \rightarrow n'$
oscillations that would allow super GKZ cosmic rays (CR) \cite{GKZ} on 
Earth, requires
that along most of the $\sim (100)$ Mega parsec distance travelled by the 
neutron,
the B field must be smaller than $10^{-15}$ Gauss. While this seems
like a severe demand, the authors of \cite{BB} argue that it is 
conceivably achieved in large extragalactic patches . 

Our purpose in this brief note is to discuss some theoretical 
implications of having a relatively ``fast'' $n-n'$ oscillation in the 
context of simple mirror models. We also note some new phenomenological
and cosmological consequences of this possibility. 

We consider two consider two classes of mirror models: one without 
supersymmetry and the other with supersymmetry. We point out that in the 
context of
non-supersymmetric models, simplest realizations of ``fast''
$n-n'$ oscillations together with a nonvanishing neutrino mass
would require the existence of light ($\sim 30-3000$ GeV) color
triplet boson. This could provide an interesting test of the
model. In the presence of supersymmetry, a new class of
models can be constructed where the ordinary and mirror bino's
$\tilde{B}$ and $\tilde{B}'$ mix to give $n-n'$ transition. These models 
lead to new phenomena
such as $K^0-{K'}^0$ as well as Higgs-mirror-Higgs mixing with
interesting implications. In this case, there exist severe constraints on 
the supersymmetry breaking parameters of the model. 

We also point out that such ``fast'' $n-n'$ oscillation times would
require that the reheat temperature after inflation has to be lower than 
about 100 GeV in order to maintain the required relative ``coolness'' of 
the mirror sector in such models. Such low reheat temperatures would 
necessitate the existence of new mechanisms for the origin of matter.

We also make a phenomenological remark that fast $n-n'$ transitions lead 
to the interesting possibility of having neutrons
  effectively "tunnel" through large amount of material. This for
instance could lead to an upward flux of neutrons from the Earth during
solar flares.

\section{Models for ``fast'' $n-n'$ oscillation and implications}
In this section, we consider minimal theoretical schemes which can
lead to ``fast'' $n-n'$ oscillations and study their consequences.
We will consider the minimally replicated standard model gauge
group i.e. $[SU(3)_c\times SU(2)_L\times U(1)_Y]\times
[SU(3)'_c\times SU(2)'_L\times U(1)'_Y]$ where the unprimed group
operates in the visible sector and the primed group operates in
the mirror sector. The two sectors communicate only via gravity as
well as via very weakly interacting particles. If the only
communications between the two sectors were via gravity, the
effective $n-n'$ operator $u^cd^cd^c\dot {u'}^c{d'}^c{d'}^c$ would
have a strength of $1/M^5_{P\ell}$ and would lead to oscillation
times which are minuscule and of no consequence. We therefore need
other interactions between the two sectors in a way that respects
gauge invariance and renormalizability in order to get sizable
$\tau_{nn'}$. We outline two such models below. Both the models
will build on the basic particle content of the mirror model given
below (for the nonsupersymmetric case):

\begin{center}
\begin{tabular}{|c||c||c|}\hline
particles & visible sector & mirror sector \\ \hline Gauge bosons
& $W, Z, \gamma$ & $W', Z', \gamma'$\\
Matter &$Q \equiv (u, d)$ & $Q'\equiv (u', d')$\\
& $u^c, d^c, e^c, \nu^c$ &  $u^{c'}, d^{c'}, e^{c'}, \nu^{c'}$\\
 Higgs fields & $H$ & $H'$ \\ \hline
\end{tabular}
\end{center}
Clearly in the above table, the fields $\nu^c$ and ${\nu'}^c$ are singlets
under both the gauge groups and mirror partners of each other. If
they are electroweak singlets, they could couple to particles in either
sector\cite{bento}. A simpler version of the theory where the sectors 
communicate
only via the term mixing the $\nu^c$ with ${\nu'}^c$ can be obtained if
there is an extra $U(1)$ in both sectors, such that both $\nu^c$ and
${\nu'}^c$ are nonsinglets under this and there is a connector field
$\phi$ with quantum numbers such that we have the coupling
$\nu^c{\nu'}^c\phi$ allowed. The vev of the $\phi$ field would then
connect the two sectors. In any case we will assume that the interaction
Lagrangian for the model to consist of two terms: ${\cal L}~={\cal L}_Y+
{\cal L}'$ where ${\cal L}_Y$
in each sector will
have the standard form with usual Yukawa
couplings as follows:
\begin{eqnarray}
{\cal L}_Y~=~h_u QH u^c+h_d Qd^c\tilde{H}+h_\ell 
Le^c\tilde{H}+h_\nu L\nu^cH\\
\nonumber + (Q\rightarrow Q', L\rightarrow L'\cdot\cdot\cdot etc)
\end{eqnarray}
The  gauge couplings and the
above Yukawa couplings do not lead to $n-n'$ oscillation and we need to
add an effective dimension 6 operator in each sector plus a connecting
term:
\begin{eqnarray}
{\cal L}'~=~\frac{1}{M^2}(\nu^cu^cd^cd^c + {\nu'}^c{u'}^c{d'}^c{d'}^c)\\ +
\nonumber \delta M\nu^c{\nu'}^c + h.c.
\end{eqnarray}
Given these interactions, having sizable $n-n'$ transition and also
nonvanishing neutrino masses puts constraints on the parameters $M$
and $\delta M$ of the model. In the case at hand we have $(B-L-B'+L')$
conserved.

In this model, the neutrino is a Dirac fermion due to $(B-L-B+L')$
conservation, with a
mass in the range of 0.1 eV connecting the $\nu$ and $\nu'$. This is given
by the usual seesaw diagram except that instead of a Majorana mass for the
right handed neutrino, we have the $\delta M$ interaction. This gives
\begin{eqnarray}
m_\nu\simeq \frac{(h_\nu v_{wk})^2 }{\delta M}
\end{eqnarray}
This implies that $h^2_\nu\frac{1}{\delta M}\simeq
10^{-14}$ GeV$^{-1}$.

The six quark operator leading to $n-n'$ oscillation has the strength:
\begin{eqnarray}
G_{6q}~\simeq \frac{1}{M^4 \delta M}\simeq
\frac{10^{-14}~GeV}{h^2_\nu M^4}
\end{eqnarray}
  $G_{6q}$ is related to the $n-n'$ transition operator roughly by
the formula $G_{n-n'}\sim\Lambda^6_{QCD} G_{6q}\sim 
10^{-4}G_{6q}$\cite{shrock} (all in GeV units).
Putting all this together and requiring that $\tau_{n-n'}\simeq 1$
sec. implies that $\sqrt{h_\nu} M \simeq 30 $ GeV. For
$h_{\nu}\simeq 1-10^{-4}$, we must have $M\simeq 30-3000$ GeV.

To see the meaning of this scale M, we can imagine a higher scale theory,
where there is a particle connecting the two quarks and the $\nu^c q$
whose exchange would lead to the interaction $u^cd^cd^c\nu^c$. This
particle would be a color triplet and must have mass $M\sim 30-3000$
GeV in order to have required ``fast'' $n-n'$ oscillation. It must couple
only to the $u^cd^d$ and $d^c\nu^c$ and not to any other fermion of the
standard model to be consistent. For example if it coupled to $u^ce^c$ it
would lead to extremely fast proton decay and will therefore be ruled
out. The existence of such a low mass particle could be searched for in
colliders and thereby provide a test of the possibility of ``fast''
$n-n'$ oscillation.

This particle will decay to two jets and will imitate the squark
in an R-parity violating supersymmetric theory. Present
experiments would eliminate any such particle below a 100 GeV.
\footnote{A simple way to avoid having such a low mass color
triplet field would be to have two standard model singlet
fermions, one responsible for neutrino mass and another for $n-n'$
oscillation. But in such a case, one will have to add extra
symmetries to explain why they do not couple to each other.}

Since the neutrino is a Dirac fermion, observation of neutrinoless
double beta decay will rule out this model. It will not lead to
any oscillation between active and sterile neutrinos and as such
cannot accommodate the LSND results. Therefore if MiniBooNe
confirms the LSND results, this model will be ruled out.
Furthermore, there is no neutron-anti-neutron oscillation in this
case.

\section{ Majorana neutrino as a way to accommodate LSND}
The model above can be modified to incorporate a Majorana neutrino
as well as the LSND results as follows. If the $(B-L-B'+L')$  symmetry
responsible for neutrino being a Dirac fermion can be broken by
adding a mass term $M_N\nu^c\nu^c$, this model leads to Majorana
neutrinos in each sector. Furthermore, if we assume that mass
parameters $M_N$ are different from those in the mirror sector,
one could make the mirror neutrino masses to be larger (say $\sim
1 $ eV) thereby making it possible to accommodate the LSND results
in the usual manner\cite{bere}. In such a version, the high scale
sector is mirror asymmetric which would go more naturally with asymmetric
inflation whereas the TeV scale low energy sector is fully mirror
symmetric, so that $n$ and $n'$ have the same mass. Let us discuss
what constraints are imposed on this model by a ``fast'' $n-n'$
oscillation.

A new feature of this model compared to the previous model is that
it will lead to oscillations between $n$ and $\bar{n}$ with a
strength  $\simeq \frac{M_N}{M^4(\delta M)^2}$ which gives
\begin{eqnarray}
\tau_{n-\bar{n}}\simeq \tau_{n-n'}\frac{\delta M}{M_N}
\end{eqnarray}
To be consistent with the present lower limit on
 $\tau_{n-\bar{n}}\leq 10^{8}$ sec.\cite{GR} with $\tau_{n-n'}\sim 1$ 
sec., we must have $M_N\leq 10^{-8} 
\delta M$ making the $\nu^c$ and $\nu^c$ a 
pseudo-Dirac pair. This is a very high degree of fine tuning of 
parameters.
One could interpret this as follows: (a) if the $n-n'$ oscillation
time scale is as fast as second and (b) if the mirror neutrinos are
responsible for explaining the LSND results and/or the neutrinos are 
established to be Majorana fermions, then $n-\bar{n}$
oscillation time should not be very much higher than its present
lower limit of $10^8$ seconds without making the level of fine tuning much 
worse. A search for $n-\bar{n}$ would then
provide a test of this model.

\section{A supersymmetric model}
The constraint of a low mass color triplet scalar particle can be avoided
if we consider the following supersymmetric model. Another way to 
look at it is to ask if the low mass scalar triplet of the previous 
sections could be a superpartner. This assumption has other constraints 
that we discuss below.

This model is based on the same particle content as in Table I except for
the fact that each field in the table is to be understood as a superfield
that contains a fermion as well as a boson in it and there will 
be two
Higgs fields $H_{u,d}$ as required in MSSM. The neutrino
masses in this model could arise from the existence of the $\nu^c$ field
as in the nonsupersymmetric case and become a Dirac fermion or they could
alternatively arise from the R-parity violating terms in each sector. The
$n-n'$ oscillation in this case arises from the presence of
 bino-mirror-bino mixing (bino being the superpartner of the $U(1)_Y$ 
gauge field of the standard model) 
 so that the neutrino mass and the $n-n'$
oscillation are decoupled from each other leading to very different 
character for this model.

The superpotential of this model is given by
\begin{eqnarray}
W~=~h_u QH_u u^c+h_d Qd^c{H_d}+h_\ell Le^c{H_d}+h_\nu L\nu^cH_u+
\\ \nonumber \lambda_q u^cd^cd^c\\
\nonumber + (Q\rightarrow Q', L\rightarrow L'\cdot\cdot\cdot etc)
\end{eqnarray}
The communication between the visible and the mirror sector is done via
the bino mixing term $M_{BB'}\tilde{B}\tilde{B}'$ as already
mentioned. This term is gauge
invariant but breaks supersymmetry softly and is allowed. A typical 
Feynman diagram giving rise to $n-n'$ oscillation is shown in Fig. 1 and 
has the
strength
\begin{eqnarray}
G_{6q}~\simeq \frac{\lambda^2_q M_{BB'}}{M^4_{\tilde{q}}M^2_B}
\end{eqnarray}
For  $M_{BB'} \leq M_B\simeq 100$ GeV,
and $M_{\tilde{q}}\simeq $ TeV, we get the constraint $\lambda^2_q 
M_{BB'}\simeq 10^{-2}$ GeV. 

In this case there is another constraint coming from the fact that a 
gluino  mediated graph can lead to $N-\bar{N}$ oscillation and present 
bounds on this process imply that 
\begin{eqnarray}
\tau_{n-n'}\simeq \frac{M^2_B}{M_{BB'}M_{\tilde{G}}} \tau_{n-\bar{n}}.
\end{eqnarray}
This implies that for $\tau_{nn'}\sim 1$ sec., the gluino has to be 
superheavy ($
M_{\tilde{G}}\sim 10^{10}$ GeV for $M_{BB'}\sim M_B\sim 100$ GeV.) This 
represents a severe degree of fine tuning and is ceratinly different from 
the conventional supersymmetry scenarios.

\begin{figure}[h]
\centering
\includegraphics{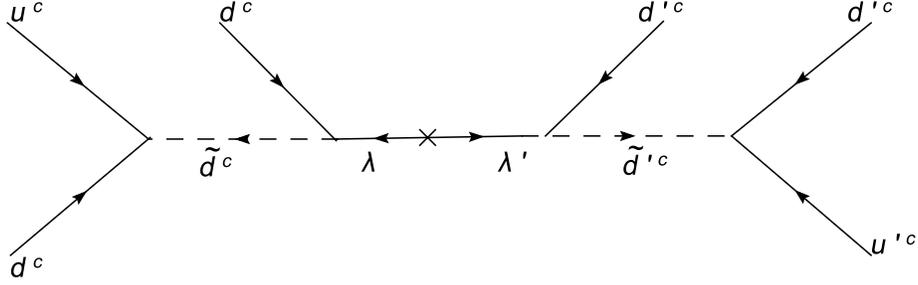}
\caption{The supersymmetric diagram for neutron mirror-neutron
oscillation caused by $\tilde{B}$ -$\tilde{B'}$ mixing.}
 \end{figure}

\begin{figure}[h]
\centering
\includegraphics{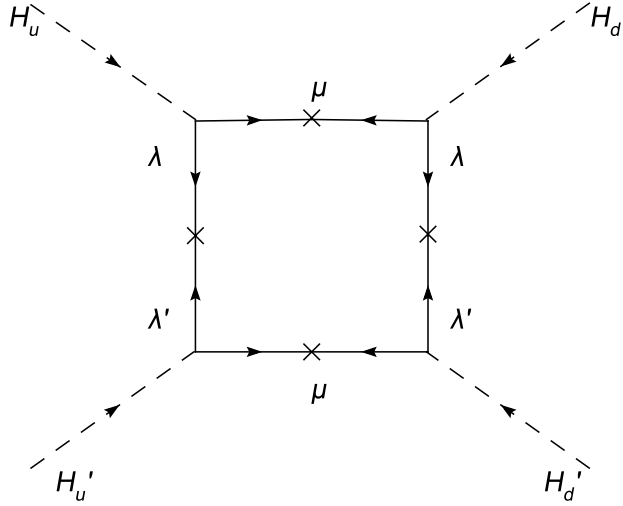}
\caption{Higgs mirror-Higgs mixing induced by $\tilde{B}$
-$\tilde{B'}$ mixing.}
 \end{figure}

If we accepted this scenario, there are cosmological implications
which put further constraints. This has to do with a
 class of new effects in the model that lead to mixing between
normal and mirror Higgs fields via radiatively induced interactions of 
type $\lambda_{H}(H_uH_d
H'_uH'_d)$. They arise at the one loop level via
diagrams of type in Fig.2. This effect can be estimated to be  
$\lambda_{eff}\simeq \frac{g^4_1 
M^2_{BB'}\mu^2}{16\pi^2 M^4_{S}}\simeq
10^{-4}$ in the simplest assumption of $M_{BB'}\simeq
\mu \simeq M_{S}\sim $ TeV. Note that this estimate is independent of the 
gluino mass. The value of 
$\lambda_{eff}$ will have implications for
cosmology and maintaining the temperature asymmetry between the
two sector\cite{dolgov} that we discuss below.

\section{Cosmology of fast $n-n'$ oscillation}
We saw from the above discussion that a transition time $\tau_{n-n'}\simeq
1$ sec. corresponds to a strength of the six quark operator of order
of $10^{-20}$ GeV$^{-5}$. This has to be consistent with another
requirement of the mirror models that the asymmetry in the temperatures
must be maintained down to the BBN epoch. To ensure this, we must have
the process $u^cd^cd^c\rightarrow {u'}^c{d'}^c{d'}^c$ induced by the
$n-n'$ operator must be out of equilibrium all the way down to the same
epoch. To see what constraint it imposes on the model, we note that the
out of equilibrium condition is given by:
\begin{eqnarray}
10^{-40}T^{11}\leq \sqrt{g^*}\frac{T^2}{M_{P\ell}}
\end{eqnarray}
This implies that above a temperature of $T_*\simeq 300$ GeV, the $n-n'$
interaction will bring the visible and the mirror sectors into
equilibrium, which is undesirable from the point of view of BBN. This
means that the reheat temperature of the universe after inflation must be
less than 300 GeV. This of course puts other constraints on the model e.g.
how to generate baryon asymmetry etc.

In the particular case of the supersymmetric model, the induced
$H^2{H'}^2$ can also lead to transitions between the visible and the
mirror sector. The strength of this transition is  $\sim 10^{-8} 
T$.
Comparing this with the Hubble expansion rate, the out of equilibrium
condition gives that for the first case $T\leq T_* \simeq 10^9$ GeV, the 
two sectors are
in equilibrium. Clearly this is in contradiction with the same constraint
from $n-n'$ transition.

This would imply that the supersymmetric model for
fast $n-n'$ transition given above is incompatible with cosmology unless 
further fine tuning of parameters is imposed ( such as vanishing $\mu$ 
terms for instance).

 \section{Phenomenological implications}

We now turn to discuss some phenomenological implications of
having a
``fast'' $n-n'$ oscillation.

First we consider the propagation of reactor neutrons. Clearly the
deficiency here is the "short length" over
which the Grenoble neutrons propagate which for these hypothermal
neutrons amounts to a $1 sec$ time interval. This is so since one
needs to strongly shield the earths magnetic field so as not to
quench to much even slow oscillations like the above $n
\rightarrow \bar{n}$ oscillations with oscillation time of $10^8
sec$. However much faster (say $sec^{-1}$) $ n \rightarrow n'$
oscillation rates of the type considered here could survive much
larger -say $ B= 10^{-3}$ Gauss fields.

Far longer propagation of
order the neutron $15$ minutes lifetime are experienced by
neutrons detected during solar flare. Travelling with mildly
relativistic speeds the $\sim 150$ million Km distance from sun
spots to earth they could readily oscillate into $n'$s in the intervening
space which have $B< ~ 10^{-3}$ Gauss. This yields a
factor two reduction in the neutron flux which however cannot be
ascertained unless we have an accurate knowledge of the neutron flux at
the source.

Note however that a satellite located at a few earth radii
distance would detect some flux of up going neutrons even when the
solar flare occurs at the opposite hemisphere and is completely
eclipsed by the earth . The point is that all the n's making up
$50\%$ of the initial flux penetrate through the earth and then
-just like in \cite{BB} super-GKZ scenario some fraction therefore
will oscillate back into neutrons. Clearly to optimize the $n'
\rightarrow n$ oscillations the satellite should be located as far
as possible ,increasing travel time and minimizing the B fields .
Unfortunately for distances $l >> R_{Earth}$ the solid angle and
the frequency and duration of the eclipsed Solar flares will be
minimal . Consider however a generic satellite which is just ~ one
earth radius where the  magnetic field ( falling like
$(R/{R_{earth}})^{-3}$ ) is $\sim 0.1$ Gauss. Along the way to it
$\sim  10^{-4}$ of the $n'$s will have reconverted back into
ordinary neutrons- and during total eclipse even such a weak
signal delayed by just $\sim  3 R_{Earth}/c \sim  0.06$ sec may be
detectable. In this short note we have not attempted to see if
there are indeed enough relevant data of this type , or in
connection with reactor,pulsed intense neutron beams which could
potentially impact on the \cite{BB} suggestion \footnote{It is
amusing to note parenthetically that the fact that $ |n>+|n'>$ and
$|n>-|n'>$ are the propagating states does not ( Like in the $K
\bar{K} - K_L K_S$  system ) prolong the life time of the
antisymmetric combination . The reason being that the $n$ and the $n'$
-unlike the kaon and anti-kaon- decay into different final states
$(pe\nu - p'e'\nu')$ and the decay amplitudes cannot (
negatively) interfere.}.

\section{Summary and conclusion}
In this brief note we have explored to what extent neutron-mirror neutron
oscillation, a phenomenon which is in principle possible in the mirror
models, can be realized in the context of realistic models. We find that
while  a ``fast'' $n-n'$ oscillation can be realized in a
nonsupersymmetric model, it imposes a nontrivial constraint on the model
of having a 10-1000 GeV color triplet scalar boson in the visible sector
which could be detected at the LHC. We also note that if $n-n'$
oscillation time is as fast as one second and LSND result is explained
by mirror neutrinos, then $n-\bar{n}$ oscillation should be observable
unless the parameters of the model are severely fine tuned. We also show
that cosmology of these
models require that the reheat temperature after inflation in these models
must be less than 300 GeV.

\section*{Acknowledgments}
\vskip -.5cm

We would like to thank Z. Berezhiani and  G. Steigman for useful 
discussions. This work is supported by National Science Foundation Grant 
No. PHY-0354401.

\end{document}